\documentstyle[twocolumn,aps,epsfig,floats]{revtex}
\begin{document}

\twocolumn[\hsize\textwidth\columnwidth\hsize\csname %
@twocolumnfalse\endcsname

\title{Proximity Effects and Quantum Dissipation in the Chains of
YBa$_2$Cu$_3$O$_{6+x}$}
\author{Dirk K. Morr and Alexander V. Balatsky}
\address{Theoretical Division, MS B262, Los Alamos National
Laboratory, Los Alamos, NM 87545}
\date{\today}
\draft
\maketitle

\begin{abstract}

We argue that the results of recent scanning tunneling microscopy,
angle-resolved photoemission and infrared spectroscopy experiments
on the CuO chains of YBa$_2$Cu$_3$O$_{6+x}$ are consistently
explained within a proximity model by the interplay of a {\it
coherent} chain-plane and an {\it incoherent} interchain coupling. We
show that the CuO$_2$ planes act as an ohmic heat bath for the
electronic degrees of freedom in the chains which induces a
substantial quantum dissipation. Below the planar $T_c$, charge
excitations in the chains acquire a {\it universal}
superconducting gap whose phase and magnitude are {\it momentum}
dependent. We predict that the magnitude of this gap  varies
non-monotonically with the hole concentration in the chains.

\end{abstract}


]

Little is known about the coupling of the CuO$_2$ planes and CuO
chains in YBa$_2$Cu$_3$O$_{6+x}$ (YBCO) and how the onset of
planar superconductivity affects the electronic degrees of freedom
in the chains. The observation of a superfluid component in the
chains below the planar $T_c$ in infrared spectroscopy (IS)
experiments by Basov {\it et al.}~\cite{Bas95} suggests that the
chains undergo a proximity-induced superconducting (PSC)
transition. This scenario is consistent with recent scanning
tunneling microscopy (STM) experiments \cite{Edw94,Loz} which
reported a large gap $\sim 15-25$ meV in the chain density of
states (CDOS) below $T_c$. Additional information comes from
angle-resolved photoemission (ARPES) experiments which observed a
sharp peak as well as a large incoherent ''hump" in the chain
spectral function, $A_c$ \cite{Sch98}.

In this Letter we show that the above-cited results of IS, STM and ARPES
experiments are consistently explained within a
proximity model by the interplay of a {\it coherent} chain-plane
(CP) and an {\it incoherent} interchain (IC) coupling \cite{com1}. We demonstrate that the
coherent/incoherent nature of these couplings causes the CuO$_2$
planes to be an ohmic heat bath for the electronic degrees of
freedom in the chains. As a result, the chains exhibit a
substantial quantum dissipation, similar to the case of a
two-level system coupled to set of harmonic oscillators
\cite{Leg87}. With the onset of planar superconductivity a
frequency gap opens in the dissipative heat bath, and a Fermi-gas
like peak appears in the chain spectral function. Simultaneously, charge excitations in the chains acquire a {\it universal}
superconducting (SC) gap whose phase and magnitude are {\it
momentum} dependent. We predict that the magnitude of this gap
varies non-monotonically with the hole
concentration in the chains. Since the intra-bilayer coupling in
YBCO was shown to be coherent \cite{bilayer}, our results
immediately imply that the c-axis transport in YBCO, a topic
extensively discussed over the last few years \cite{c-axis}, is
necessarily dominated by coherent electronic hopping.

For the calculation of the CDOS and spectral function, measured in
surface sensitive ARPES and STM experiments, we consider an
infinite array of CuO-chains which is aligned along the x-axis and
coupled to a single CuO$_2$-plane via a hopping term $t_\perp$
\cite{com2}. We neglect the second CuO$_2$-plane in the bilayer
unit cell since the planes are only weakly coupled \cite{bilayer}.
We assume that the superconducting pairing interaction resides
solely in the plane, and describe the planar electrons by a
mean-field BCS-hamiltonian. The hamiltonian for the chain-plane
system is thus given by
\begin{eqnarray}
{\cal H} &=&  \sum_{k, \sigma} \epsilon_{\bf k}
c^\dagger_{k,\sigma} c_{k,\sigma} + \sum_k \left( \Delta_k
c^\dagger_{k,\uparrow} c^\dagger_{k,\downarrow} + h.c. \right)
\nonumber \\
 & & \quad  + \sum_{k, \sigma} \zeta_{\bf k} d^\dagger_{k,\sigma}
d_{k,\sigma} - t_\perp \sum_{k, \sigma} \left( c^\dagger_{k,\sigma}
d_{k,\sigma} + h.c. \right)
\label{Hpc}
\end{eqnarray}
where $c^\dagger_k, d^\dagger_k$ are the fermionic creation
operators in the plane and chain, respectively. The tight-binding
dispersions for the plane and chain electrons, $\epsilon_{\bf k}$
and $ \zeta_{\bf k}$, are given by
\begin{eqnarray}
\epsilon_{\bf k} &=& -2t_p \Big( \cos(k_x) + \cos(k_y) \Big)
 -4t_p^\prime \cos(k_x) \cos(k_y)  -\mu_p \ , \nonumber \\
\zeta_{\bf k} &=& -2 t_c \cos(k_x) -\mu_c \ .
\label{dispersion}
\end{eqnarray}
Here, $t_p, t_p^\prime (t_c)$ are the planar (chain) hopping
elements between  nearest and  next-nearest neighbors,
respectively, and $\mu_p (\mu_c)$ is the planar (chain) chemical
potential. The planar SC gap is given by $\Delta_{\bf
k}=\Delta_{SC} \ (\cos(k_x) - \cos(k_y))/2$.

An effective action for the chain electrons, $S_c$, is
obtained by integrating out the planar electronic degrees of
freedom. One obtains
\begin{equation}
{\cal S} = \sum_{\omega_n,l,m} \int dk_x \, {\hat d}^+_{l}\ {\hat
G}_c^{-1}(\omega_n,k_x,l,m) \ {\hat d}_{m} \ ,
\label{c_action}
\end{equation}
where $l,m$ are chain indices, and
\begin{equation}
{\hat d}^+_{l} = \left( d^*_{l,\uparrow}(k_x,\omega_n),
d_{l,\downarrow}(-k_x,-\omega_n) \right) \ .
\end{equation}
For the matrix ${\hat G}_c^{-1}$ one has
\begin{equation}
{\hat G}_c^{-1}=
\left(
\begin{array}{cc}
G_0^{-1}(k_x,\omega_n) & F_0^{-1}(k_x,\omega_n) \\
F_0^{-1}(k_x,\omega_n) & - G_0^{-1}(-k_x,-\omega_n)
\end{array}
\right)   \, ,
\label{Ginv}
\end{equation}
where
\begin{eqnarray}
G_0^{-1} &=& \ \left( \, i\omega_n - \zeta_{\bf k} \, \right) \ \delta_{l,m}
\nonumber \\
& & \hspace{-0.5cm} -t_\perp^2 \, N^{-1}  \sum_{k_y}  \,  {i \omega_n +
\epsilon_{\bf k} \over (i \omega_n)^2 -
(\epsilon_{\bf k})^2 - \Delta_k^2 } \
 e^{ ik_y(l-m) }  \, ; \nonumber  \\
F_0^{-1}&=& -t_\perp^2 \, N^{-1}  \sum_{k_y}
 {\Delta_k \over (i \omega_n)^2 -
(\epsilon_{\bf k})^2 - \Delta_k^2 } \ e^{ ik_y(l-m) } \, .
\label{G0F0}
\end{eqnarray}
The chain Greens function for coherent IC coupling is
straightforwardly computed from Eqs.(\ref{c_action}) and
(\ref{Ginv}). In contrast, in the case of incoherent IC coupling
where the coherent correlations between the chains are destroyed,
one evaluates Eq.(\ref{c_action}) using the ansatz
\begin{equation}
{\hat G}_c^{-1}(k_x,l,m,\omega_n) = {\hat G}_c^{-1}(k_x,\omega_n)
\ \delta_{l,m} \, .
\label{decoup}
\end{equation}

Our theoretical results for the CDOS and spectral function
presented in the following are obtained by using a set of band
parameters, $t_p=300$ meV, $t^\prime_p/t_p = -0.4, \mu_p/t_p =
-1.18$, extracted from ARPES experiments on YBCO \cite{Sch98} and
a planar SC gap, $\Delta_{SC} \approx 35-40$ meV, representative
of that in slightly underdoped YBCO compounds \cite{Mag96}. The
chain Fermi momentum, $k^c_F \approx 0.25 \pi$, was extracted from
INS \cite{Mook96} and STM experiments \cite{Edw94}, and yields
$\mu_c/t_c=-1.41$, corresponding to a hole concentration of $x
\approx 50 \%$ in the chains.

\begin{figure} [t]
\begin{center}
\leavevmode
\epsfxsize=7.5cm
\epsffile{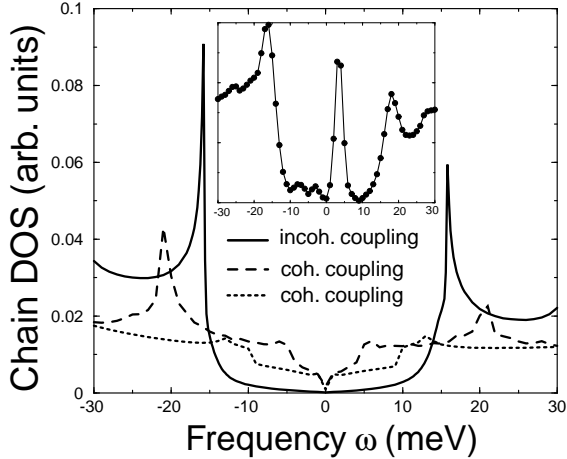}
\end{center}
\caption{Chain DOS in the SC state for $t_\perp=0.4t_p$ and
$\Delta_{SC}=40$ meV. Solid line: incoherent IC coupling. Dashed
line: coherent IC coupling. Dotted line: coherent IC coupling with
$\Delta_{SC}=100$ meV. Inset: Experimental data taken from
Ref.~\protect\cite{Loz}. }
\label{dos1}
\end{figure}
In Fig.~\ref{dos1} we present our theoretical results for the
chain DOS, $N(\omega) = N^{-1} \sum_{k}  A_c(k, \omega)$
\cite{com3}, together with the experimental results of
Ref.~\cite{Loz}. The possible impurity origin of the peak in the
experimental CDOS at $\omega \approx 3$ meV \cite{Lpc} is beyond
the scope of the present letter and will be addressed in a future
study. Our theoretical result for the chain DOS for incoherent IC
coupling (solid line) are in good agreement with the
experimentally measured magnitude of the chain gap,
$\Delta_{c}^{exp} \approx 17$ meV, as well as the observed
particle-hole asymmetry of the CDOS.  The simultaneous appearance
of an anomalous chain Greens function, $F_0$, in Eq.(\ref{Ginv})
demonstrates that the induced gap is related to the onset of
superconductivity in the chains. In contrast, the CDOS for
coherent IC coupling (dashed line) and the same set of parameters
does not only possess a much smaller gap \cite{Atkpc}, but its
overall frequency dependence is also inconsistent with the
experimental data. We explored various sets of band parameters and
did not find an improved agreement with the experimental data,
even for an unphysically large superconducting gap of
$\Delta_{SC}=100$ meV (dotted line in Fig.~\ref{dos1}). Moreover,
for incoherent CP coupling (and thus incoherent IC coupling), no
gap is induced in the CDOS and $F^{-1}_0 \equiv 0$ in
Eq.(\ref{Ginv}); hence no superconducting correlations are present
in the chains, in disagreement with the results of IS experiments
\cite{Bas95}. We thus conclude that the experimental STM and IS
data provide strong evidence for an incoherent interchain and
coherent chain-plane coupling in YBCO.

\begin{figure} [t]
\begin{center}
\leavevmode
\epsfxsize=7.5cm
\epsffile{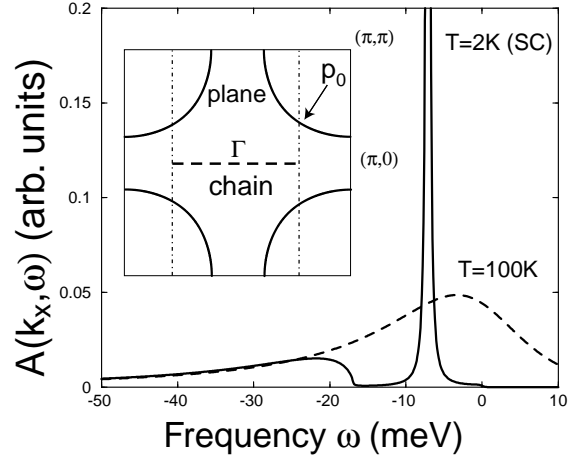}
\end{center}
\caption{Frequency dependence of $A_c$ for incoherent IC coupling
at the chain Fermi momentum, $k^c_F=0.25 \pi$, in the normal
(dashed line) and superconducting state (solid line).} \label{ak0}
\end{figure}
Before we can address the physical origin of the different CDOS
for coherent and incoherent IC coupling, it is necessary to study
the spectral function, $A_c( k, \omega)$, for incoherent IC
coupling, in more detail. In Fig.~\ref{ak0} we present our
numerical results for $A_c$ at $k^c_F$ \cite{com3}; for comparison
with ARPES measurements we have multiplied $A_c$ with the Fermi
distribution function $n_F=(\exp(\omega/k_BT)+1)^{-1}$, which
shifts the peak in $A_c$ to lower energies. In the normal state
(dashed line) the spectral function exhibits a broad peak,
indicating a strong quasiparticle damping. The source of the
quasiparticle dissipation is best understood by analytically
evaluating the integrals on the r.h.s.~of Eq.(\ref{G0F0}). One
obtains for the retarded chain Greens function at momentum $k_x$
\begin{equation}
G^R_c(k_x, \omega) = {1 \over \omega - \zeta_k + i \alpha} \ ,
\label{Akapprox0}
\end{equation}
where $\alpha= t_\perp^2 / {\bar v}$, ${\bar v} = \left(
\partial \epsilon_{\bf k} / \partial k_y \right)|_{{\bf p}_0} $,  and ${\bf
p}_0$ is a momentum on the planar Fermi surface (FS) with
$p_0^x=k_x$ (see inset of Fig.~\ref{ak0}). Due to the coupling,
$t_\perp$,  the chain electrons can decay into a continuum of
planar states perpendicular to the chain direction. The planes
thus represent a dissipative environment which gives rise to a
frequency independent quasiparticle damping, $\alpha$. This form
of dissipation is analogous to the ``ohmic case" discussed by
Leggett {\it et al.}~\cite{Leg87} for a two-level system  coupled
to a set of one-dimensional harmonic oscillators, which constitute
a similar dissipative ``heat bath". Thus, $\alpha$ does {\it not}
arise from a Coulomb-type interaction, but from the different
dimensionality of the chain and plane system.

With the onset of superconductivity, the planar heat bath acquires
a gap and the frequency dependence of $A_c$ (solid line in
Fig.~\ref{ak0}) changes dramatically. At low frequencies, the
damping of the chain electrons is strongly reduced and a Fermi-gas-like low-frequency quasiparticle peak (LFQP) appears. The
suppression of the quasiparticle damping arises from a kinematic
constraint since a chain electron can only decay into the
continuum of planar states if its frequency exceeds the necessary
energy, ${\bar \Delta}(k_x)=|\Delta({\bf p}_0)|$, to break a
Cooper-pair. Hence, one recovers a substantial quasiparticle
damping only for $\omega > {\bar \Delta}(k_x)$. Moreover, the LFQP
is shifted downward in energy from the position of the broad peak
at $\omega=0$ in the normal state (cf.~solid line), indicating the
presence of a gap for electronic excitations in the chains. This
gap arises from a virtual hopping of a chain electron into the
planar continuum states, a process which lowers the electron's
ground state energy; hence, one expects the gap to be of order
$t_\perp^2/v_F$. Moreover, an analysis of $F_0$, Eq.(\ref{G0F0}),
shows that the phase of the induced gap is {\it momentum}
dependent and changes at $k_n$, where ${\bf K}_{node}=(k_n,k_n)$
is the momentum of the planar SC nodes. The separation of the LFQP
and the incoherent background by a dip in Fig.~\ref{ak0} is
reminiscent of planar ARPES data \cite{Chu98} where a decrease of
the fermionic damping in the SC state due to the opening of a spin
gap gives also rise to a ``peak-dip-hump" structure in the
spectral function.

In order to better understand the momentum dependence of $A_c$ below $T_c$, it is necessary to analytically study the
retarded chain Greens function, $G^R_c$, in Eq.(\ref{G0F0}). One
obtains
\begin{equation}
G^R_c = B(\omega,k_x)^{-1} \ \Big[ \omega + \zeta_k +  { \alpha
\omega \over \sqrt{{\bar \Delta}^2 - \omega^2} }  \Big] \ ,
\label{Gapprox1}
\end{equation}
where
\begin{eqnarray}
\label{G0inv}
B(\omega,k_x) &=&  \omega^2 \left(1 +  { \alpha \over \sqrt{ {\bar
\Delta}^2 - \omega^2 } } \right)^2 \nonumber \\ & & \quad  -
\left( \zeta_k \right)^2 - \left( { \alpha {\bar \Delta} \over
\sqrt{ {\bar \Delta}^2 - \omega^2 } }\right)^2 \ ,
\end{eqnarray}
and $\omega=\omega+i\delta$. The dispersion, $\chi_k$, of the LFQP
is determined by Re$\,B(\omega=\chi_k) = 0$. In the limit
$\alpha^2+\zeta_k^2 \ll {\bar \Delta}^2 $, i.e., close to the
chain Fermi points, one has
\begin{equation}
\chi_k= \pm \sqrt{\alpha+ \zeta_k^2} \ , \label{lfgap}
\end{equation}
where $\alpha$ is the {\it universal} SC chain gap, which is
independent of the planar gap, $\Delta_{SC}$. In the opposite
limit, $\alpha^2+\zeta_k^2 \gg {\bar \Delta}^2$, one obtains
\begin{equation}
\chi_k =  \pm {\bar \Delta}  \left[ 1 - { 2 \alpha^2 {\bar
\Delta}^2 \over (\alpha^2 + \zeta_k^2 )^2 } \right] \ ,
\label{hfgap}
\end{equation}
and the LFQP is thus confined to frequencies $|\omega|<|{\bar
\Delta}|$ for all $k_x$. In addition, one finds a high-frequency
quasi-particle peak (HFQP) whose dispersion is given by $ \phi_k =
\pm \sqrt{ \zeta_k^2 + \alpha^2 }$. Since $\phi_k > |{\bar
\Delta}|$, the HFQP is overdamped with
\begin{equation}
A_c(k, \omega=\phi_k) = \sqrt{\phi_k^2-{\bar \Delta}^2}\ {\phi_k +
\zeta_k \over \alpha \phi_k^2  }  \sim \alpha^{-1} \ .
\label{Akhf}
\end{equation}
In the limit $\zeta_k \gg \max\{ {\bar \Delta}, \alpha\}$, the
HFQP follows the dispersion of the free chain electrons, $\phi_k =
\zeta_k$, with a {\it momentum independent} intensity,
$A_c=\alpha^{-1}$.

\begin{figure} [t]
\begin{center}
\leavevmode
\epsfxsize=7.5cm
\epsffile{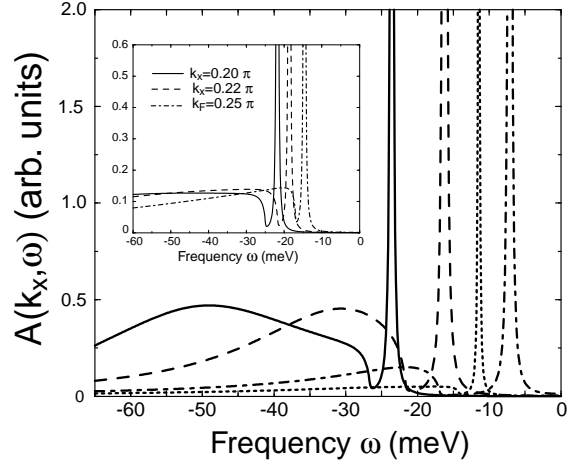}
\end{center}
\caption{$A_c$ as a function of frequency for $t_\perp=0.18t$,
$\mu_c=-1.41t_c$ and four momenta: $k_x=0.19\pi$ (solid line),
$k_x=0.22\pi$ (dashed line),  $k^c_F=0.25 \pi$ (dashed-dotted
line), and $k_x=0.27\pi$ (dotted line). Inset: $A_c$ for
$t_\perp=0.40t_p$ and three momenta.} \label{ak1}
\end{figure}

In Fig.~\ref{ak1} we present our numerical results for $A_c$ in
the SC state for several momenta around $k_F^c$.
For momenta much below $k^c_F$ ($k_x=0.19 \pi$, solid line;
$k_x=0.22 \pi$, dashed line) $A_c$ possesses a HFQP and a LFQP,
which both move to higher energies as $k_x$ approaches $k^c_F$.
The dispersion of the LFQP is much weaker than that of the HFQP,
since the LFQP is located close to $\bar{\Delta}(k_x)$ which
changes only weakly with momentum while the HFQP's dispersion is
given by $\zeta_k$. Note the momentum independence of the HFQP's
intensity. At the Fermi momentum, $k^c_F=0.25 \pi$, the HFQP has
disappeared and only a broad incoherent background remains with a
maximum close to its onset frequency ${\bar \Delta}\approx -17$
meV. For momenta above $k^c_F$ ($k_x=0.27 \pi$, dashed-dotted
line) the LFQP shifts again to lower energies, while its intensity
and that of the incoherent background decreases rapidly. An
increased CP coupling ($t_\perp=0.4t_p$, see inset of
Fig.~\ref{ak1}) leaves $A_c$ qualitatively unchanged, albeit with
a much broader HFQP. The momentum and frequency dependence of
$A_c$ shown in Fig.~\ref{ak1} is in qualitative agreement with
recent ARPES experiments by Schabel {\it el al.}~\cite{Sch98} in
the SC state of YBa$_2$Cu$_3$O$_{6.95}$. They also report a broad
dispersing peak for momenta $k_x \ll k^c_F$, while for $k_x
\approx k^c_F$, $A_c$ exhibits a hump and a sharper quasi-particle
peak (cf.~Fig.~7d in Ref.~\cite{Sch98}).

\begin{figure} [t]
\begin{center}
\leavevmode
\epsfxsize=7.5cm
\epsffile{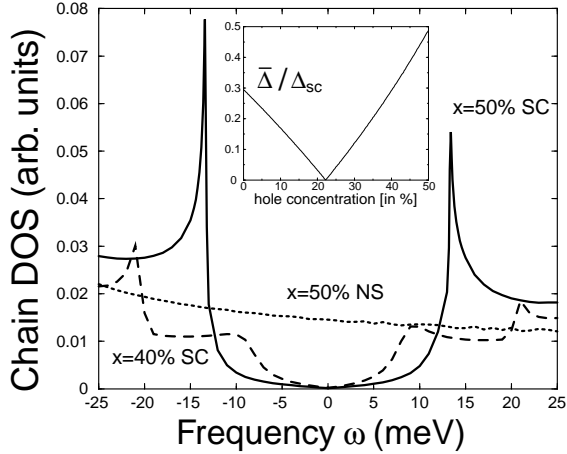}
\end{center}
\caption{DOS for $t_\perp=0.4t_p$, and  two different hole
concentration, $x$: {\it (a)} $x=50\%$, SC state (solid line) and
normal state (dotted line), {\it (a)} $x=40\%$ SC state (dashed
line). Inset: $\bar{\Delta}(k_F^c)/\Delta_{SC}$ as a function of
hole concentration in the chains.}
\label{dos_comp}
\end{figure}

The gap induced in the CDOS for incoherent IC coupling varies
strongly with the hole concentration, $x$, in the chains, as shown
in Fig.~\ref{dos_comp}.
For $t_\perp=0.4t_p$, one finds $\alpha > {\bar \Delta}(k^c_F)$
and hence the frequency of the gap edges in the CDOS is determined
by ${\bar \Delta}(k^c_F)$ (cf. Eq(\ref{hfgap})).  Since ${\bar
\Delta}(k^c_F)$ decreases as the hole concentration is reduced
from $x=50\%$ to $x=40\%$ (see inset of Fig.~\ref{dos_comp}), the
magnitude of the induced gap also decreases, in agreement with our
results in Fig.~\ref{dos_comp}. The gap in the CDOS is the
smallest for $x=22$\%, since here $k^c_F=k_n$ and ${\bar
\Delta}(k^c_F)=0$. We thus predict that the gap in the CDOS varies
non-monotonically with $x$: as $x$ is decreased from $x=50\%$ the
induced gap first decreases, exhibits a minimum at $x=22\%$ and
then increases again for even smaller hole concentrations.
Moreover, for small frequencies we find $N(\omega) \sim \omega$
which is a direct consequence of the linear momentum dependence of
$\Delta_k$ in the vicinity of the nodes. In contrast, the chain
DOS in the normal state (dashed line) is only weakly frequency and
doping dependent.

The origin of the qualitatively different CDOS for coherent and
incoherent IC coupling is twofold. First, for coherent IC
coupling, the chain states are described by a two-dimensional
momentum, ${\bf k}$, and the dimensionality of the {\it coherent}
chain array and the CuO$_2$ planes is the same. Thus, the CuO$_2$
planes do not constitute a dissipative environment for the
electronic degrees of freedom in the chains. Second, the changes
which occur in the CDOS for  a given frequency, $\omega$, upon
entering the SC state arise from those electronic states which can
reduce their ground state energy by a virtual hopping process into
the planes. It follows from Eqs.(\ref{Ginv}) and (\ref{G0F0}) that
in the case of coherent IC coupling these states must satisfy
$t_\perp^2 \gg \omega^2-\epsilon_k^2-\Delta_k^2$. Simple phase
space counting shows that the number of states which undergo a
shift in the ground state energy is much smaller for coherent than
for incoherent IC coupling. As a result, the CDOS for incoherent
IC coupling exhibits a much larger SC gap than that for coherent
IC coupling.

Two explanations for the absence of a coherent IC coupling in YBCO
are possible. First, STM experiments have provided evidence for
inhomogeneous doping in the chains. The resulting
distribution of Fermi momenta likely prevents the coherent
coupling between momenta in the vicinity of $k^c_F$,
and thus renders the coupling incoherent. Second, a series of
experimental probes have reported charge and spin inhomogeneities
(stripes) in the CuO$_2$ planes which are aligned parallel to the
chains \cite{Mook00}. Since planar electrons are likely scattered
by these inhomogeneities, a coherent coupling only exists between
those chains which can be connected by an electron path that is
not intersected by a stripe. This scenario can be tested in the
underdoped YBCO compounds where the stripe-stripe distance
increases and coherent IC coupling between an increasing number of
chains is expected.

Recently, there has been an extensive discussion on the relative
strength of incoherent versus coherent hopping of carriers between
the planes \cite{c-axis}. Since the CP hopping is a part
of the hopping process between the bilayers in YBCO, our results
support the notion that coherent hopping is a crucial part of the
communication of electrons between the planes. Though the CP
hopping, $t_\perp$, in the bulk is likely somewhat reduced from
the value extracted above (the distance between the surface chain
layer and the next CuO$_2$ layer is smaller than the corresponding
separation in the bulk \cite{Lpc}) we expect it to remain
coherent.

Finally, the phase of the induced SC gap as well as its
non-monotonic doping dependence, Fig.~\ref{dos_comp}, are salient
features of the PSC state which distinguish it from other possible
ground states, e.g., a charge-density-wave. We thus propose
phase-sensitive Josephson tunneling or Josephson STM experiments
as a crucial test for our above scenario.

In summary, we propose a scenario in which the results of recent
IS,  STM and ARPES experiments on the chains of YBCO can
consistently be explained by the interplay of a coherent
chain-plane coupling and an incoherent interchain coupling. We
show that the CuO$_2$ planes act as a dissipative environment for
the electronic degrees of freedom in the chains and induce a
substantial quantum dissipation. We find that the chains exhibit
superconducting correlations and a {\it universal} gap below
$T_c$, and predict that the magnitude of this gap varies
non-monotonically with the hole concentration in the chains.

We would like to thank P.W. Anderson, W.A. Atkinson, A. Chubukov,
J.C. Davis (JCD), A. de Lozanne (AdL), J. Schmalian, and J. Zaanen
for stimulating discussions, and JCD and AdL for providing us with
their experimental data prior to publication. This work has been
supported by the Department of Energy at Los Alamos.


\begin{thebibliography}{99}
\bibitem{com1} Since no direct IC coupling is possible in YBCO, any IC coupling is necessarily mediated by the CuO$_2$ planes.
\bibitem{com2} The surface layer in cleaved YBCO is the chain layer,
followed by a CuO$_2$ plane.
\bibitem{com3} Our numerical results for $A_c=-2{\rm
Im}G^R_c/\pi$ are obtained from Eq.\protect(\ref{G0F0}) via the
analytical continuation $i\omega_n \rightarrow \omega+ i\, \delta$
with $\delta=0.1$ meV.
\end{thebibliography}
\end{document}